\documentclass[sigconf]{acmart}
\AtBeginDocument{%
  }

\usepackage[caption=false,font=normalsize,labelfont=sf,textfont=sf]{subfig}
\usepackage{makecell}
\usepackage{listings}
\usepackage{svg}
\usepackage{color}
\usepackage{float}
\usepackage{multirow}
\usepackage{xcolor}
\usepackage{fancybox}
\usepackage{enumitem}
\usepackage{colortbl}
\usepackage{xspace}
\hypersetup{
    colorlinks=true,
    linkcolor=black,
    anchorcolor=black,
    filecolor=black,      
    citecolor=black,
    urlcolor=black
}

\newcommand{\xzp}[1]{{{#1}}}
\newcommand{\tool}{\textsc{\textbf{TypeUp}}\xspace}

\newcommand{\find}[1]{%
  \vspace{0.5em}%
  \noindent%
  \fcolorbox{black}{gray!10}{
    \parbox{\dimexpr\linewidth-2\fboxrule-2\fboxsep}{\em #1}%
  }%
  \vspace{0.5em}%
}

\author{Zhipeng Xue}
\orcid{0000-0002-9060-4064}
\affiliation{%
  \institution{The State Key Laboratory of Blockchain and Data Security, Zhejiang University}
   \city{Hangzhou}
  \country{China}
}
\email{zhipengxue@zju.edu.cn}

\author{Zhipeng Gao}
\orcid{0000-0003-3030-9917}
\authornote{Corresponding Authors}
\affiliation{%
  \institution{Shanghai Institute for Advanced Study, Zhejiang University}
   \city{Shanghai}
  \country{China}
}
\email{zhipeng.gao@zju.edu.cn}

\author{Xing Hu}
\orcid{0000-0003-0093-3292}
\affiliation{
  \institution{The State Key Laboratory of Blockchain and Data Security, Zhejiang University}
   \city{Hangzhou}
  \country{China}
}
\email{xinghu@zju.edu.cn}

\author{Jingyuan Chen}
\orcid{0009-0005-7481-3332}
\affiliation{%
  \institution{College of Education, Zhejiang University}
   \city{Hangzhou}
  \country{China}
}
\email{jingyuanchen@zju.edu.cn}

\author{Xin Xia}
\orcid{0000-0002-6302-3256}
\authornotemark[1]
\affiliation{
  \institution{The State Key Laboratory of Blockchain and Data Security, Zhejiang University}
   \city{Hangzhou}
  \country{China}
}
\email{xin.xia@acm.org}

\author{Shanping Li}
\orcid{0000-0003-2615-9792}
\affiliation{%
  \institution{The State Key Laboratory of Blockchain and Data Security, Zhejiang University}
   \city{Hangzhou}
  \country{China}
}
\email{shan@zju.edu.cn}

\copyrightyear{2026}
\acmYear{2026}
\setcopyright{rightsretained}
\acmConference[ICSE '26]{2026 IEEE/ACM 48th International Conference on Software Engineering}{April 12--18, 2026}{Rio de Janeiro, Brazil}
\acmBooktitle{2026 IEEE/ACM 48th International Conference on Software Engineering (ICSE '26), April 12--18, 2026, Rio de Janeiro, Brazil}
\acmPrice{}
\acmDOI{10.1145/3744916.3764565}
\acmISBN{979-8-4007-2025-3/26/04}

\begin{document}

\title{Automating Just-In-Time Python Type Annotation Updating}

\begin{abstract}
Type annotations are more and more popular in Python projects to avoid type errors caused by Python's dynamic typing feature.
However, when developers change source code, these type annotations are often neglected or overlooked, resulting in outdated and inconsistent type annotations. 
Such obsolete type annotations can hinder program comprehension, mislead developers, and even introduce bugs in the future. 
Therefore, it is necessary to avoid and correct these inconsistent type annotations from the very beginning.
In this work, we argue that obsolete type annotations can be reduced and even avoided by automatically updating type annotations alongside code changes. 
We refer to this task as “Just-In-Time (JIT) type annotation updating”. To solve this task, we propose a novel LLM-based approach named \tool (\underline{Type} Annotation \underline{Up}dator) to automate this task. \tool can automatically generate new type annotations based on the old type annotations and corresponding code changes.
Specifically, \tool guides LLM to perform type annotation updates by eliciting its knowledge and logical reasoning power and learning from similar code changes. 
The evaluation results show that \tool outperforms the state-of-the-art type inference approach (i.e., TypeGen) by 41.9\% on our task. 
Moreover, we conducted a practical application with real-world software projects, 20 out of 25 type annotation updates generated by our approach have already been confirmed by developers, showing our approach’s practical value in real-world environments.

\end{abstract}



\maketitle

\section{Introduction}
Python has become increasingly popular among developers for its usage convenience and extensive third-party libraries. 
As a dynamically typed programming language, Python determines the type of a variable at runtime, rather than at compile time. 
In other words, developers do not need to specify a variable type when they declare it. 
This distinctive feature makes Python extremely flexible and easy to use~\cite{yang2022complex}. 
However, \textit{every coin has two sides,} this feature also brings risks and threats to Python software projects because Python type errors can only be found when relevant code is executed, not when the code is compiled. 
Researchers find that 30\% of the issues or questions raised by developers at GitHub or Stack Overflow are related to Python type errors~\cite{khan2021empirical}. 
To reduce the introduction of potential type errors, nowadays more and more Python software projects have adopted type annotations as a way to document expected types of different elements (e.g., function arguments, return values, variable assignments)~\cite{jin2021start}.

\begin{figure}
    \centering
    \includegraphics[width=0.45\textwidth]{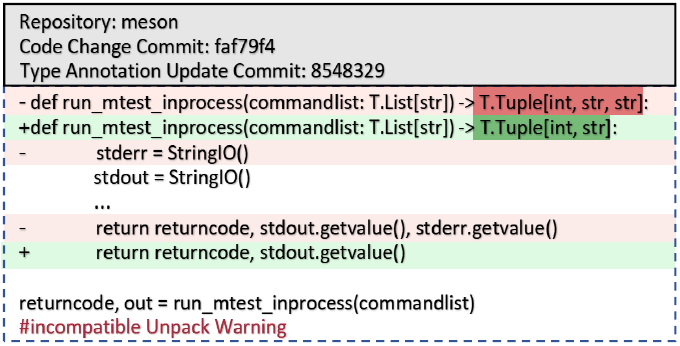}
    \caption{An example of outdated type annotation}
    \label{fig:motiexample}
    \vspace{-20pt}
\end{figure}

Based on the value of type annotations, recent studies highlighted that type annotations are manually maintained and often edited alongside code changes~\cite{di2022evolution, yu2019characterizing, ore2021empirical}.
However, developers can easily forget or ignore the updates of such type annotations when changing source code, which results in outdated and inconsistent type annotations. 
Figure~\ref{fig:motiexample} presents an example of inconsistent type annotation in meson project~\cite{meson}. 
The original function \texttt{run\_mtest\_inprocess} returned a tuple value of three elements, namely \texttt{T.Tuple[int, str, str]}. 
The developer later removed the last element, \texttt{stderr.getvalue()}, from the returned values. 
However, what he/she neglected was that the type annotation of the return values was not updated correspondingly. 
After updating, only two elements were returned by the function and the type annotation became obsolete and incorrect (e.g., \texttt{T.Tuple[int, str, str]} should be corrected as \texttt{T.Tuple[int, str]}). 
Before being fixed by developers, this obsolete type annotation had existed for over two years until successfully updated by our approach. 
During this time, it may hinder program comprehension, complicate code reviews, and even mislead developers who perform the subsequent development. 

Once the obsolete type annotations are introduced, they become incorrect and unreliable documentation of ever-decreasing accuracy, which can mislead developers or even introduce bugs in the future~\cite{tobin2017migratory, spiza2014type}.  
Currently, a number of studies already explored how to infer and predict correct type annotations for Python element~\cite{mir2022type4py, peng2022static, peng2023generative}, however, how to update Python type annotations based on code changes has never been investigated. 
Inspired by the example in Figure~\ref{fig:motiexample}, it is desirable to have a tool to automatically update Python type annotations with code changes, which can possibly reduce or even avoid obsolete type annotations.

\begin{figure}
    \centering
    \includegraphics[width=0.45\textwidth]{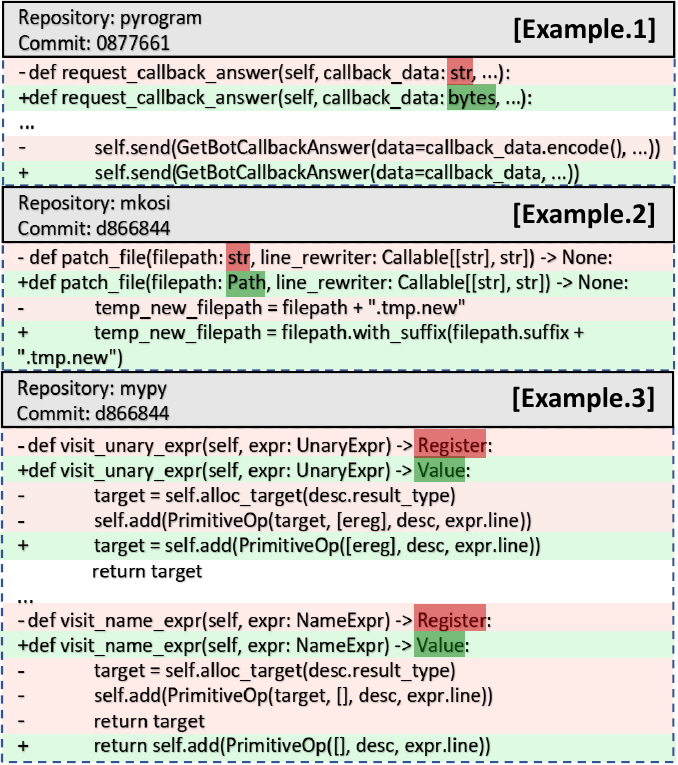}
    \caption{Examples of correctly updated type annotations}
    \label{fig:introexample}
\end{figure}

In this work, we propose a novel task of updating Python type annotations with code changes, namely ``Just-In-Time Type Annotation Updating''. 
However, this newly proposed task is non-trivial in terms of the following aspects: 
\begin{itemize}[leftmargin=*]
    \item \textbf{Lacking available datasets:} Currently, there are no available datasets regarding the Python type annotation updates. 
    Constructing the dataset of type annotation evolution alongside code changes is the first step to building the tool and estimating the model performance. 
    \item \textbf{Inferring correct type annotation from source code:} 
    It is often difficult to determine if the current type annotation is valid or not by just reading the source code. 
    A more reliable way is to check the code change history regarding type annotations and infer the correct type annotation based on code change information. 
    An example is shown in Figure~\ref{fig:introexample} Ex.1, the type of \texttt{callback\_data} is hard to determine based on the changed code contexts. However, one can easily infer it as \texttt{bytes} since the argument \texttt{data} previously was assigned as \texttt{callback\_data.encode()}. 
    \item \textbf{Capturing the semantic correlation between type annotation and code change:}  
    The code change can affect the type annotation in a subtle and intricate way. 
    As shown in Figure~\ref{fig:introexample} Ex.2, the developer refactored the function argument \texttt{filepath}, resulting in its type changing from \texttt{str} to \texttt{Path}. 
    How to capture the semantic relationship between the code change and type change is another challenge of this study. 
    \item \textbf{Learning project-specific context:} 
    Sometimes checking the code changes alone is not sufficient. 
    Figure~\ref{fig:introexample} Ex.3 presents such a case, even though the code change was presented, one can not easily infer the updated type (e.g., \texttt{Value} is a project-specific type) due to his/her unfamiliarity with the project context. 
    Under such circumstances, a similar update in code change history can provide additional clues to fill this gap. 
    For example, a similar update involving the invocation of the function \texttt{PrimitiveOp} suggests the return type annotation should be changed from \texttt{Register} to \texttt{Value}.
\end{itemize}

To address the aforementioned challenges, we first collected the first dataset for this JIT type annotation updating task from 450 GitHub repositories. 
We then propose a novel approach, named \tool (\underline{Type} Annotation \underline{Up}dator), to automate this task, which can automatically generate new type annotations based on the old type annotations and corresponding code changes. 
Different from type inference studies~\cite{allamanis2020typilus, yan2023dlinfer, mir2022type4py} that generate the target type annotation from source code, our tool infers the correct type annotations from code changes, which provides additional information about \textit{why} and \textit{what} type annotation should be updated (to). 
To capture the semantic relationship between the code change and type annotation co-change, we employ Large Language Models (LLMs) agents in this study due to their great potential in code comprehension and program understanding~\cite{hou2024large, xue2024selfpico, dai2025less}. 
Finally, to enhance our model with project-specific knowledge, we adopt the Retrieval Augmented Generation (RAG) technique to retrieve similar code changes in project development history and provide \tool with project-specific context (e.g., user-defined types). 
The key idea of \tool is to guide LLM to perform type annotation updates by eliciting its knowledge and logical reasoning power and learning from similar code changes. 
Overall, \tool consists of the following three phases: 
\begin{itemize}[leftmargin=*]
    \item \textit{Knowledge Database Building:} As there is no existing dataset for type annotation updating, at this stage, we collect more than 36K examples from 450 GitHub projects to construct a knowledge database. 
    Each example contains a code change and a type co-change pair. 
    Following that, we adopt Coditt5~\cite{zhang2022coditt5}, a pre-trained model for code edits, to convert each code change to its vector representation, so that similar code changes can be easily retrieved by calculating similarity scores between two vectors. 
    \item \textit{Augmented Information Retrieving:} After building our database, we perform information retrieval to augment LLMs with project-specific knowledge. 
    Given a new code change, two kinds of information are retrieved for subsequent type updating, i.e., similar guiding examples and type updating candidates (e.g., project user-defined types). 
    Similar guiding examples are used to elicit LLM's logical reasoning power and type updating candidates are provided to narrow down type selections. 
    \item \textit{Type Annotation Updating:} In this stage, our tool employs the RAG technique to guide LLM to perform type updating. 
    In particular, we first use LLM to reason about the retrieved guiding examples, eliciting LLM's logical reasoning power on ``\textit{why the code change results in this type update?}''.
    Following that, we query LLM with the current code change and potential type candidates, enhanced by its logical reasoning ability to capture the semantic relationship between code change and type co-change. As a result, \tool can correctly choose the most likely type from type candidates and perform updating action. 
\end{itemize}

To evaluate the effectiveness of \tool, we randomly selected 500 code-annotation co-change instances from our constructed dataset as the testing set, with the remaining instances serving as the knowledge database. 
Evaluation results show that \tool successfully predicts 359 correct updated type annotations, surpassing the state-of-the-art type inference model (i.e., TypeGen) by 41.9\%. 
Moreover, we conducted a practical application with real-world GitHub projects. 
\tool fixed 25 outdated type annotations in 10 GitHub projects, and we submitted these type annotations updates to GitHub developers. 
Notably, 20 of these updates were confirmed as correct by the developers, demonstrating the practical usability of our tool. 
In summary, this paper makes the following contributions: 
\begin{enumerate}[leftmargin=*]
    \item We present the first study on Python type annotation updating. We build a dataset with 36,796 code-type co-change samples. 
    To the best of our knowledge, this is the first dataset on this research topic. 
    \item We propose a novel task, Just-In-Time Type annotation updating, in this paper, and we develop a tool, named \tool, to automate this task. 
    \tool is based on LLMs and introduces customized RAG techniques to effectively handle the characteristics of this task. 
    \item We extensively evaluate \tool using real-world projects in GitHub. \tool is shown to outperform several baselines by a large margin. 
    Moreover, 20 out of 25 type annotation updates generated by our approach have already been confirmed by developers, showing our approach's potential for increasing software quality and maintainability. 
    \item We have released our replication package~\cite{replication}, including source code and dataset to facilitate the follow-up works.  
\end{enumerate}
\section{Problem and Usage Scenario}

\subsection{Problem Formulation}
This work targets at automating JIT type annotation updating for Python, i.e., automatically updating type annotations with code changes. 
This task can be formalized as follows: given the pre-change and post-change version of a code snippet \textbf{\textit{c}}, \textbf{\textit{c'}}, the corresponding pre-change and post-change version of the type annotation are \textbf{\textit{t}} and \textbf{\textit{t'}} $(\textbf{\textit{t}} \neq  \textbf{\textit{t'}})$, our task aims to find a function \textit{f}, so that $f(\textbf{\textit{c}}, \textbf{\textit{c'}}, \textbf{\textit{t}}) = \textbf{\textit{t'}}$. 
We refer to \textbf{\textit{c}}, \textbf{\textit{c'}}, \textbf{\textit{t}}, \textbf{\textit{t'}} as \textit{old code}, \textit{new code}, \textit{old type} and \textit{new type} respectively. 
In this study, we focus on updating annotations of parameter types and return types, which are mostly relevant to developers~\cite{jin2021start, di2022evolution} in software development.


\subsection{Usage Scenario}
Our work presents a tool, named \tool, that takes a code change and its old type annotation as input, aiming at generating the newly updated type annotations. 
The usage scenario of our tool is as follows: Our tool can be used to assist developers in performing Just-In-Time type updating. 
When developers make code changes, \tool can automatically provide type update suggestions based on LLM's reasoning ability and project development histories. 
As the example shows in Figure~\ref{fig:motiexample}, when the developer removed an element from the return values, he/she may easily forget to update its type annotation. 
With the help of our tool, the developer can quickly confirm this type updating through one click, increasing the consistency between the code and its type and alleviating the error-prone code reviewing process. 

Our tool can also be incorporated with Python type-checking tools (e.g., mypy~\cite{mypy}, Pyre~\cite{pyre}) to fix existing type annotation errors. 
\xzp{As shown in Section~\ref{sec:wild}. 
By leveraging Pyre, we collected 25 incompatibilities caused by outdated type annotations that were overlooked after code changes across 10 popular GitHub projects. These issues remained unresolved until we submitted pull requests to address them. With the help of \tool, 20 out of 25 outdated type annotations have already been confirmed by developers, showing our approach’s potential for increasing software quality and maintainability.}


\section{Approach}
\begin{figure}
    \centering
    \includegraphics[width=0.5\textwidth]{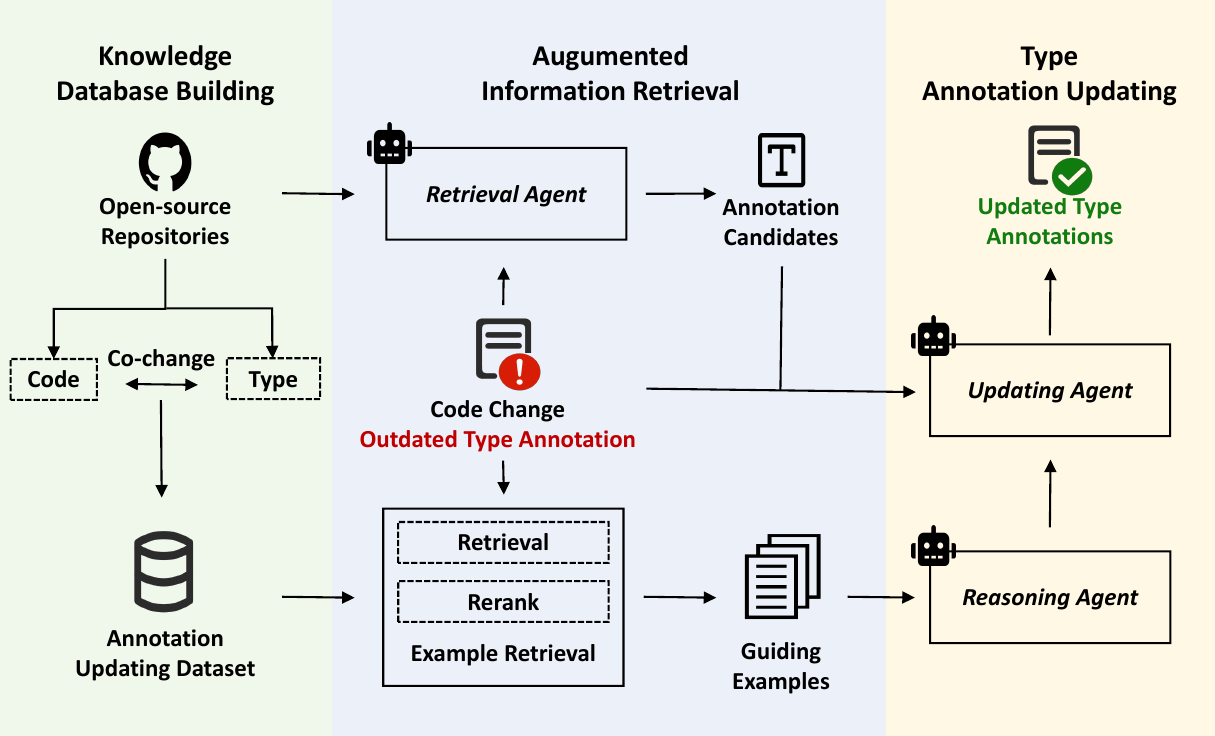}
    \caption{Overview of \tool}
    \label{fig:overview}
\end{figure}

The overall framework of our approach is illustrated in Figure~\ref{fig:overview}. 
It consists of three stages, knowledge database building, augmented information retrieving, and annotation type updating. 
Specifically, we first extracted code-type co-change examples from GitHub repositories as our database. 
Then, we retrieve augmented information (i.e., similar guiding examples and type-updating candidates) from our knowledge database. 
Finally, given a code change and its associated old type annotation, we query LLM with the above augmented information. The LLM is guided by similar examples to reason about the semantic correlation between the code change and type co-change, and automatically generates a new type annotation (from candidate pool) to replace the old one. 
Below, we present the details of our approach.

\subsection{Knowledge Database Building}
\label{sec.dataset}
In this phase, we extract code changes and type co-changes from the project commit history and encode code changes into vector representations so that they can be easily retrieved.
\subsubsection{Dataset Preparation}
We build our knowledge database based on the type annotation evaluation dataset proposed by Di Grazia et al~\cite{di2022evolution}, which contains 61,861 type annotation change-related commits from 668 popular GitHub repositories. 
In this study, we further process their dataset for our JIT type updating task, extracting the code pre-change \textit{\textbf{c}} and code post-change \textit{\textbf{c'}}, as well as the type pre-change \textbf{\textit{t}} and type post-change \textit{\textbf{t'}} from each commit. 
In particular, for each type annotation change-related commit, we first check out the modified source code files (i.e., files with \texttt{.py} extension) of this commit. 
Then for each modified file, we identified and collected type annotations of all return values and function arguments based on parsing the source code into an abstract syntax tree (AST). 
For each type annotation, we compared its pre-change version (i.e., \textit{\textbf{t}}) and post-change version (i.e., \textit{\textbf{t'}}) and only retained changed type annotations (i.e., $\textit{\textbf{t}} \neq \textit{\textbf{t'}}$). 
We then extracted the function-level source code in both the pre-change version (i.e., \textit{\textbf{c}}) and the post-change version (i.e., \textit{\textbf{c'}}) and filtered out instances where source code remains unchanged. 
Each pre/post-change code and its pre/post-change type annotation is regarded as a data sample. 
Finally, we collected 36,796 samples from 450 GitHub repositories to construct our dataset. 
More formally, our constructed dataset is $D = \{\langle c_{i}, c'_{i}, t_{i}, t'_{i} \rangle_{i=1}^{N}\}$, where $N$ is the total number of collected samples.

\subsubsection{Database Construction}
After building our dataset, we constructed a searching database based on our dataset. 
Particularly, we adopt Coditt5~\cite{zhang2022coditt5} to encode each sample's code change (including the pre-change and post-change versions of code) into a vector representation. 
The CoditT5 is a pre-trained language model trained for code editing and transformation tasks, making it effectively captures the semantic intent behind code edits and well-suited for our type annotation update task.
By querying this database, similar code changes can be easily retrieved by calculating similarity scores between two vectors.

\begin{figure*}
    \centering
    \includegraphics[width=1.0\textwidth]{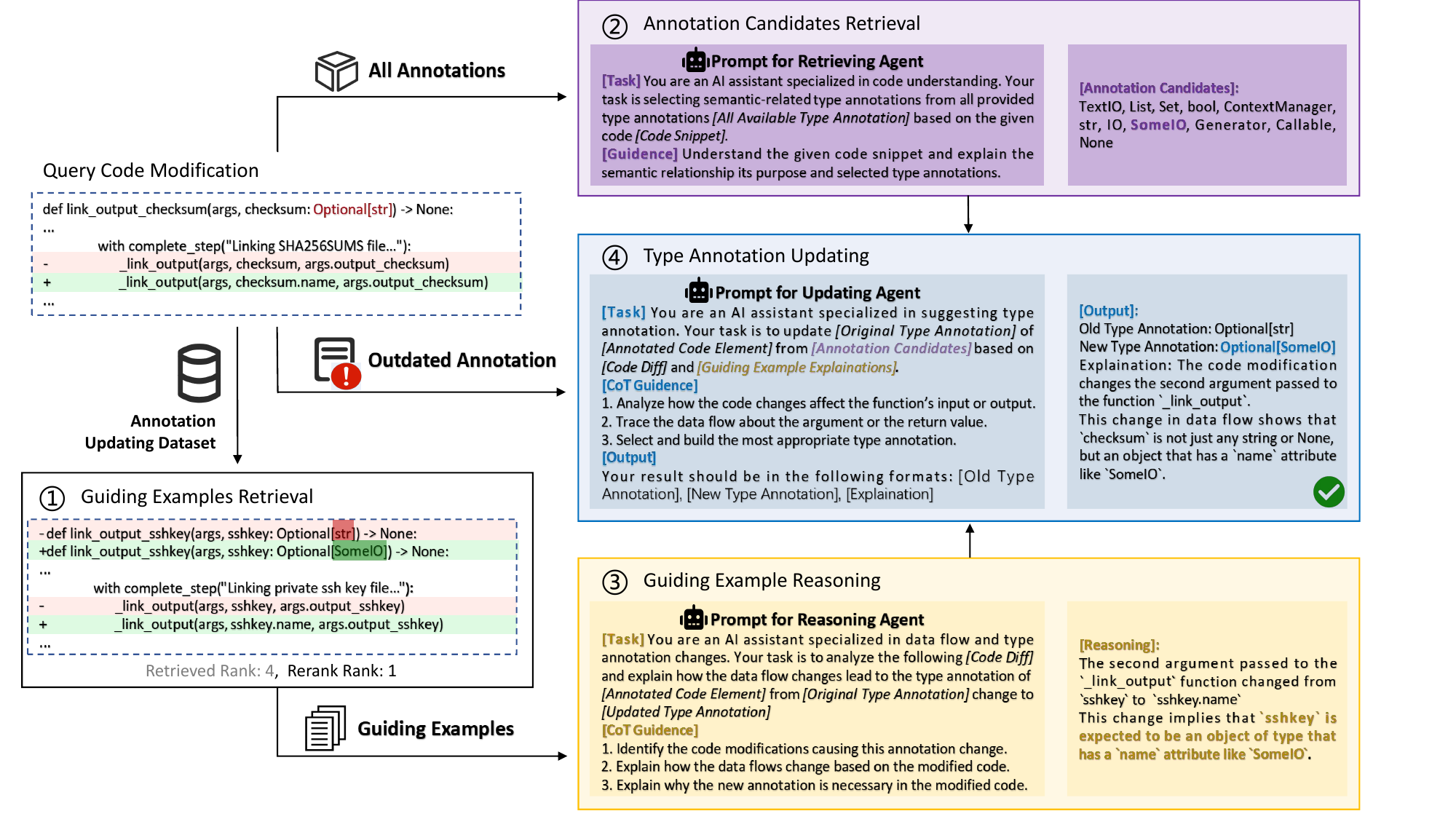}
    \caption{An example of type annotation updating}
    \label{fig:pipeline}
\end{figure*}

\subsection{Augmented Information Retrieval}
In this phase, we fetch two types of augmented information, i.e., guiding examples and annotation candidates, to guide LLM to perform subsequent type updating. 

\subsubsection{Guiding Examples Retrieval}
Since LLMs are not specifically designed or trained to comprehend code change and type annotations, in this study, we aim to elicit LLM's reasoning abilities by providing a few similar guiding samples. 
Specifically, we fetch the guiding examples as follows: (1) \textbf{Coarse-grained retrieving:} Given a new code change, we employ Coditt5~\cite{zhang2022coditt5} to encode this code change into a vector in the same way of constructing the database. 
This vector representation captures semantic-level information, enabling the retrieval of semantically similar code changes from historical commits, rather than relying solely on structural and lexical matching.
Following that, we calculate the cosine similarity scores between the code change vector and each vector in the knowledge database and retrieve top-ranked samples. 
In this preliminary study, we fetch the Top-10 most similar samples (including the similar code-change and type co-change) as our coarse-grained retrieving results. 
(2) \textbf{Fine-grained reranking:} To choose the most relevant guiding examples, we further rerank the top-10 fetched samples with a time-aware reranking strategy. Specifically, based on the coarse-grained retrieving results, we rerank them according to the commit time of each example, prioritizing those temporally closer to the query code change, i.e., from most recent to oldest.
This design reflects the intuition that more recent changes are likely more relevant in guiding the current update.
As shown in Figure~\ref{fig:pipeline}, the 4th ranked sample is reranked to the first place after the reranking process. 
Finally, we select the Top-k$(1 \leq k \leq 10)$ samples after reranking as our guiding examples. 


\subsubsection{Annotation Candidates Retrieval}
Unlike Python elementary types (e.g., \textit{str}, \textit{int}) or generic types (e.g., \textit{Union}, \textit{List}), which are widely used in most Python projects, user-defined types are created by developers themselves and/or imported from third-party libraries. 
These user-defined types are designed according to specific project contexts, making it rather difficult for LLM to generate without any project-specific knowledge. 
Therefore, it is reasonable to provide LLM with possible user-defined types as references. 
To address this context information gap, we create an annotation candidates pool in this step to include all likely user-defined type annotations. 
\xzp{Given that every class and type can serve as a potential annotation candidate, the number of possible type annotations in the project is large. It is essential to filter out candidates unrelated to code changes to prevent the LLM from being overwhelmed with redundant information.}
\xzp{To do so, we first collect all the available annotation candidates by AST. Then we leverage a \textbf{\textsc{Retrieval Agent}} by prompting LLM to identify type annotations semantic-related to the modified function.}
\xzp{
In particular, we provide a detailed task description (e.g., \textit{Your task is selecting semantic-related type annotations from all provided type annotations [All Available Type Annotation] based on the given code [Code Snippet].}) to perform this task. \textit{[All Available Type Annotation]} and \textit{[Code Snippet]} refer to the type annotation retrieved by AST and post-change code, respectively. Besides, we also provide comprehensive guidance for \textbf{\textsc{Retrieval Agent}} (e.g., Understand the given code snippet and explain the semantic relationship its purpose, and selected type annotations)}.
As a result, all potential semantic-related annotation types are extracted to make an annotation candidates pool. 
\xzp{As shown in Figure~\ref{fig:pipeline}, based on the \texttt{link\_output\_checksum} function, the elementary and generic types (e.g., \texttt{bool}, \texttt{str}, \texttt{Generator}) as well as user-defined types (e.g., \texttt{SomeIO}, \texttt{TextIO}) within this specific project are identified to make the annotation candidates pool.}

\subsection{Type Annotation Updating}
In this phase, we provide LLM with the above augmented information (i.e., annotation candidates and guiding examples) and guide LLM to reason about the semantic relationship between code-type co-change and finally generate a new type annotation to replace the old one.

\subsubsection{Guiding Example Reasoning}
Considering LLMs are not specifically designed for code changes and type annotations, it may still be difficult for LLMs to understand the semantic relationship regarding the code-type co-changes.
Therefore, it is necessary to elicit its logical reasoning power on this task by providing a few guiding examples. 
To handle this issue, we leverage a \textsc{\textbf{Reasoning Agent}} to teach LLM how to reason about ``\textit{why the code change results in a type co-change?}''.
Particularly, as shown in Figure~\ref{fig:pipeline}, the LLM is provided with detailed task descriptions (e.g., \textit{``Your task is to analyze the following [Code Diff] and explain how the type annotations of [Annotated Code Element] from [Original Type Annotation] change to [Updated Type Annotation]''}), the placeholder \textit{[Code Diff]}, \textit{[Original Type Annotation]} and \textit{[Updated Type Annotation]} denotes the code modifications, old type annotation and new type annotations of each guiding example respectively. 
Moreover, we design the chain-of-thought (CoT) reasoning prompt to guide our \textsc{\textbf{Reasoning Agent}} to reason code-type co-changes in a way similar to developers, e.g., identifying critical code changes, analyzing subsequent data flows, and providing necessary explanations. 
The CoT prompt guides our agent in thinking step-by-step like a human developer and finally elicits its logical reasoning power for this task. 
As shown in Figure~\ref{fig:pipeline}, \textsc{\textbf{Reasoning Agent}} successfully infers \texttt{sshkey} should be an object such as \texttt{SomeIO} when the function \texttt{\_link\_output} argument changes. 

\subsubsection{Type Annotation Updating}
After the LLMs gain logical reasoning power on code-type co-changes, we perform the target task of type annotation updating. 
Similar to guiding example reasoning, we leverage an \textsc{\textbf{Updating Agent}} to guide LLM on how to generate new type annotations from the current code change. 
Specifically, besides the \textit{[Code Diff]}, \textit{[Annotated Code Element]} and \textit{[Original Type Annotation]} used in \textsc{\textbf{Reasoning Agent}}, we provide \textsc{\textbf{Updating Agent}} with additional \textit{[Annotation Candidates]} and \textit{[Guiding Example Explanations]}. 
We prompt LLMs to update \textit{[Original Type Annotation]} from \textit{[Annotation Candidates]} based on \textit{[Code Diff]} and \textit{[Guiding Example Explanations]}. 
The \textit{[Annotation Candidates]} (from \textsc{\textbf{Retrieving Agent}}) provides the project-specific annotation candidates for updating, while the \textit{[Guiding Example Explanations]} (from \textsc{\textbf{Reasoning Agent}}) provides reasoning ability to capture the current code change (i.e., \textit{[Code Diff]}) with each updating candidate and generate the updated type annotation.
Similar to \textsc{\textbf{Reasoning Agent}}, we also design the CoT prompt for \textsc{\textbf{Updating Agent}} to guide type annotation updating and restrict output to follow the expected format (i.e., \texttt{[Old Type Annotation]} \texttt{[New Type Annotation]} \texttt{[Explanation]}).  
As shown in Figure~\ref{fig:pipeline}, the \textsc{\textbf{Updating Agent}} successfully generates the correct type annotation, i.e., \texttt{Optional[SomeIO]} (by combining \texttt{Optional} and \texttt{SomeIO} from annotation candidates) based on the augmented information.

\subsection{Implementation}
We implemented \tool's three agents (i.e., \textsc{\textbf{Retrieval Agent}}, \textsc{\textbf{Reasoning Agent}} and \textsc{\textbf{Updating Agent}} ) using OpenAI's GPT-3.5-turbo model as a base. 
GPT-3.5-turbo is currently one of the most powerful large language models in terms of its versatility and performance. 
For encoding the coding change into vector representations,  we utilize the CoditT5 model~\cite{zhang2022coditt5}, which is pre-trained for code edit-related tasks based on CodeT5~\cite{wang2021codet5}, demonstrating strong performance in representing code modifications.
We also adopt Chroma~\cite{Chroma}, i.e., the fastest embedding database, to make a search engine, so the similar code changes can be effectively and efficiently retrieved.  
All the details of \tool can be found at our replication package~\cite{replication}.

\section{Evaluation}
In this study, to evaluate the performance of \tool, we aim to answer the following research questions: 
\begin{itemize}[leftmargin = *]
    \item \textbf{RQ1.} How effective is \tool in JIT Python type annotation updating?
    \item \textbf{RQ2.} How effectively do different LLM-based agents contribute to the overall performance? 
    \item \textbf{RQ3.} How effective is \tool under different number of retrieved examples settings?
\end{itemize}

\subsection{Dataset Overview}
We describe the details of constructing our dataset in Section~\ref{sec.dataset}. 
In summary, we extracted 36,796 samples of code-type co-change from 450 GitHub repositories. 
In this study, we randomly sampled 500 samples of code-type co-change to serve as our test set. 
The remaining 36,296 samples were used to build the knowledge database for just-in-time type annotation updating. 
It is worth noting that, to avoid data leakage, we impose a strict temporal constraint in our study: for any testing code change, only similar code changes that occurred earlier (based on commit timestamps) are eligible for retrieval.
This ensures that the model only accesses information available at the time of prediction, thereby faithfully reflecting a realistic just-in-time type updating scenario.

To better position our dataset in the context of existing resources, we compare it with the ManyTypes4Py dataset~\cite{mt4py2021}, which is the most widely used benchmark in type inference research~\cite{peng2023generative, guo2024generating}.
While ManyTypes4Py is well suited for type inference, it is less appropriate for our focus on type updating.
In the ManyTypes4Py benchmark, most types have never been updated, making it unsuitable for training or evaluating on the Python type updating. Moreover, ManyTypes4Py only provides static type annotations along with their corresponding code context (e.g., source code files), but it does not include information about how type annotations evolve over time. In contrast, our type updating task collects type annotation changes (before and after updating) and their corresponding code changes. 

To further highlight the differences, Table~\ref{tab:dataset} demonstrates the distribution of different kinds of type annotations in ManyType4Py dataset and ours. 
Elementary indicates Python built-in types (e.g., \texttt{int}, \texttt{str}, \texttt{bool}). 
Generic indicates the combination of elementary types (e.g., \texttt{Union[str, int]}). 
User-defined indicates the user created types or third-party library imported types. 
From the table, we can see that our dataset differs from ManyTypes4Py significantly. 
More than half of the types in ManyTypes4Py are elementary types, while more than 70\% of the types in our dataset are user-defined types. 
This also points out a key difference between the type inference task and the type updating task. 
Type inference task is more concerned with predicting elementary Python types, while regarding type updating task, developers update user-defined types most often. 
As a result, the type inference tools may not shed light on this novel type updating task.  



\begin{table}
  \centering
  \caption{Statistics of Datasets}
  \label{tab:dataset}
  \begin{tabular}{lccc}
    \toprule
    Dataset & Elementary & Generic & User-defined\\
    \midrule
    ManyTypes4Py & \textbf{52.7\%} & 26.8\% &  19.8\% \\
    Ours & 16.3\% & 13.1\% & \textbf{70.6\%}\\
    \bottomrule
  \end{tabular}
\end{table}

\subsection{Baselines}
Currently, there are no methods specifically designed for our type annotation updating task. 
We adopt four type inference tools as baselines for our comparison purposes. 
\textbf{Notably, all these baselines infer type annotations from source code instead of code changes.}
\begin{itemize} [leftmargin=*]
    \item \textbf{pytype}~\cite{pytype} is a popular static type checking and inference tool developed by Google. 
    It infers the absent type annotations and checks the correctness of the program based on type matching.
    \item \textbf{Type4py}~\cite{mir2022type4py} is a deep learning based type inference tool. 
    It considers identifiers, code context, and visible type hints as features for learning to predict types. We leveraged their provided pre-trained model for this study. 
    \item \textbf{TypeGen}~\cite{peng2023generative} is an LLM-based type inference tool, which is the state-of-the-art type inference method. 
    TypeGen infers the type annotation by combining the Chain-of-Thought prompting and static domain knowledge. 
    For a fair comparison, we implement TypeGen based on GPT-3.5-turbo, the same with our settings. 
    \item \textbf{HiTyper}~\cite{peng2022static} is a hybrid type inference tool that combines machine learning models and static analysis. It validates predictions from models by static rules. HiTyer's performance heavily depends on the ML prediction model, we implement HiTyper on top of TypeGen, which is the SOTA model for type inference.  
    \item \textbf{Tiger}~\cite{wang2024tiger} is a two-stage generating-then-ranking framework to infer type annotation by fine-tuning pre-trained code models. We leverage Tiger following the default setting in its open-source repository.
\end{itemize}

\subsection{Evaluation Metrics}
We adopt three commonly used evaluation metrics~\cite{peng2022static, guo2024generating, peng2023generative} to evaluate the performance of \tool and other baselines for this JIT type annotation updating task:

    \noindent \textbf{Correct Annotations} is defined as the number of type predictions generated by an approach that exactly matches ground truth type annotations.
    \noindent \textbf{Recall} represents the proportion of all type annotation queries that are successfully predicted. It can be calculated by 
    \[
\text{Recall} = \frac{Predicted\_Annotations}{All\_Annotations\_Queries}
\]
    \noindent \textbf{Precision (Exact Match)} represents the proportion of type predictions generated by an approach that exactly matches ground truth type annotations among all the predicted types. It can be calculated by 
    \[
\text{Precision(EM)} = \frac{Exactly\_Match\_Annotations}{Predicted\_Annotations}
\]
    \noindent \textbf{Precision (Parametric Match)} represents the proportion of type predictions generated by an approach that only matches the outermost type with ground truth type annotations among all the predicted types. It can be calculated by 
    \[
\text{Precision(PM)} = \frac{Parametrically\_Match\_Annotations}{Predicted\_Annotations}
\]

For example, \textit{List[int]} and \textit{List[str]} are considered parametrically match but not exactly match since they are different types while sharing the same outmost type \textit{List}. 
During evaluation, for each tested code-type co-change sample, the updated type annotation is treated as the ground truth, which is submitted by developers in practice.
This data-driven approach ensures the ground truth labels accurately reflect real-world type updates by developers in actual development workflows.



\subsection{RQ1: Effectiveness of \tool}
In this RQ, we evaluate \tool and the aforementioned baseline methods in terms of Precision (EM), Precision (PM), Recall, and Correct Annotations. The evaluation result is illustrated in Table~\ref{tab:overall effective}, we can see that: 

\begin{itemize} [leftmargin=*]
    \item \textbf{\tool outperforms all baselines by a large margin in terms of most evaluation metrics.} 
    Specifically, compared with the state-of-the-art type inference tool, TypeGen, our tool shows an 11.6\% improvement in Precision (EM), a 14.0\% improvement in Precision (PM), and a 27.2\% improvement in Recall. 
    Furthermore, \tool successfully generates 359 type annotations that match those manually updated annotations, whereas TypeGen only predicts 253 correct updates, demonstrating the advantage of our tool on this type annotation updating task.  
    \item Tiger performs best regarding the recall metric since its re-ranking module ensures a type annotation output for any source code. However, its generation module struggles to infer the correct type annotation, which further impacts its overall performance.
    \item Type4py achieves the worst performance regarding the Precision (EM) and Precision (PM). Type4py was designed and trained to predict common element types, failing to infer the user-defined types that are predominant in our codebases.  
    \item Our approach has its advantage over static analysis rule-based methods (e.g., pytype and Hityper), this is because rule-based methods often suffer from low recall problems caused by dynamic features and external calls. 
\end{itemize}

To better analyze the effectiveness of \tool, we further conduct a type-wise evaluation with TypeGen, which outperforms other baselines in previous evaluation. 
Table~\ref{tab:type effective} shows the performance of \tool in updating three kinds of type annotations, we can see that \tool achieves better performance than TypeGen in generating three kinds of type annotations. Specifically, \tool improves the number of correct annotations by 51.2\% for elementary types, 88.1\% for generic types, and 28.0\% for user-defined types, respectively. 
We attribute this to the following reasons: (1) Instead of inferring the type annotation from the source code, our approach predicts the new type annotations based on code changes, suggesting that the code change contains valuable information for type annotation updating (e.g., data-flow changes, addition/removal of function arguments). (2) We augment LLM with project-specific context information (annotation candidates) and similar code-type co-changes (guiding examples), guiding LLM to infer updated type annotations in a way similar to developers.

\find{\textbf{Answer to RQ1:} Our approach is highly effective for JIT type annotation updating and outperforms the type inference tools by
a large margin.}


\begin{table} 
  \centering
  \caption{Overall Effectiveness Evaluation}
    \resizebox{0.47\textwidth}{!}{
  \label{tab:overall effective}
  \begin{tabular}{lcccc}
    \toprule
    Approach & Precision (EM) & Precision (PM) & Recall & \makecell[c]{Correct\\Annotations}\\
    \midrule
    pytype & 0.600 & 0.727 &  0.110 &  33\\
    Type4py & 0.071 & 0.125 & 0.754 & 27 \\
    Hityper & 0.256 & 0.455 & 0.374  & 48 \\
    Tiger & 0.356 & 0.506 & \textbf{1.00} & 178\\
    TypeGen & 0.682 & 0.736 & 0.742 & 253\\
    \tool & \textbf{0.761}& \textbf{0.839} & 0.944 & \textbf{359}\\
    \bottomrule
  \end{tabular}
  }
\end{table}

\begin{table}
  \centering
  \caption{Type-wise Effectiveness Evaluation}
  \label{tab:type effective}
  \resizebox{0.47\textwidth}{!}{
  \begin{tabular}{lccccc}
    \toprule
    Type & Approach & Precision (EM) & Precision (PM) & Recall & \makecell[c]{Correct\\Annotations}\\
    \midrule
    Elementary & TypeGen & 0.605 & 0.704 &0.732 & 43 \\
    Type & \tool & 0.714 & 0.835 & 0.938 &65 \\
    \midrule
    Generic & TypeGen & 0.525 & 0.625 & 0.672 & 42 \\
    Type & \tool & 0.711 & 0.838 & 0.933 & 79 \\
    \midrule
    User-defined & TypeGen & 0.763 & 0.786 & 0.774 & 168 \\
    Type & \tool & 0.796 &0.841  &0.951 & 215\\
    \bottomrule
  \end{tabular}
  }
\end{table}


\subsection{RQ2: Ablation Study}
The key to this task is to effectively teach LLM how to update type annotations based on code changes. 
To do so, we designed three LLM-based agents, i.e., \textsc{\textbf{Retrieval Agent}}, \textsc{\textbf{Reasoning Agent}} and \textsc{\textbf{Updating Agent}} to perform this task. 
In this RQ, we perform an ablation study to evaluate whether \textsc{\textbf{Retrieval Agent}} and \textsc{\textbf{Reasoning Agent}} provide helpful information to the \textsc{\textbf{Updating Agent}}. 
Particularly, we removed \textsc{\textbf{Retrieval Agent}}, \textsc{\textbf{Reasoning Agent}} and both from \tool respectively, denoted as \tool-RET, \tool-REA, and \tool-BOTH. 
The evaluation results are illustrated in Table~\ref{tab:ablation}, we have the following observations: 

\begin{table}
  \centering
  \caption{Ablation Evaluation}
  \label{tab:ablation}
    \resizebox{0.47\textwidth}{!}{
  \begin{tabular}{lcccc}
    \toprule
    Approach & Precision (EM) & Precision (PM) & Recall& \makecell[c]{Correct\\Annotations}\\
    \midrule
    \tool-BOTH & 0.701 & 0.777 & 0.864 & 303\\
    \tool-REA &0.713  & 0.790 & 0.920 & 328\\
    \tool-RET & 0.747 & 0.816 & 0.894 &334 \\
    \tool & \textbf{0.761}& \textbf{0.839} & \textbf{0.944} & \textbf{359}\\
    \bottomrule
  \end{tabular}
  }
\end{table}

\begin{itemize}[leftmargin=*]
    \item No matter which agent we dropped, it hurts the overall performance of our approach. 
    This shows the importance and necessity of these three LLM-based agents. 
    \item By comparing \tool-BOTH with \tool-REA and \tool-RET, it is clear that adding the \textsc{\textbf{Reasoning Agent}} and \textsc{\textbf{Retrieval Agent}} improves the overall performance. We attribute this to the ability of \textsc{\textbf{Reasoning Agent}} for LLMs' eliciting logical reasoning from guiding examples and \textsc{\textbf{Retrieval Agent}} for adding project-specific knowledge. 
    \item It is notable that even if we just use the {\textbf{Updating Agent}} (i.e., \tool-BOTH), it still achieves comparable or better performance than other baselines, which further confirms the advantage of \tool for using LLMs to learn from code changes. 
\end{itemize}

\find{
{\bf Answer to RQ2:} All three agents are effective and helpful in enhancing the performance of our model.
}

\subsection{RQ3: Sensitivity Analysis}
\begin{figure}
    \centering
    \includegraphics[width=0.45\textwidth]{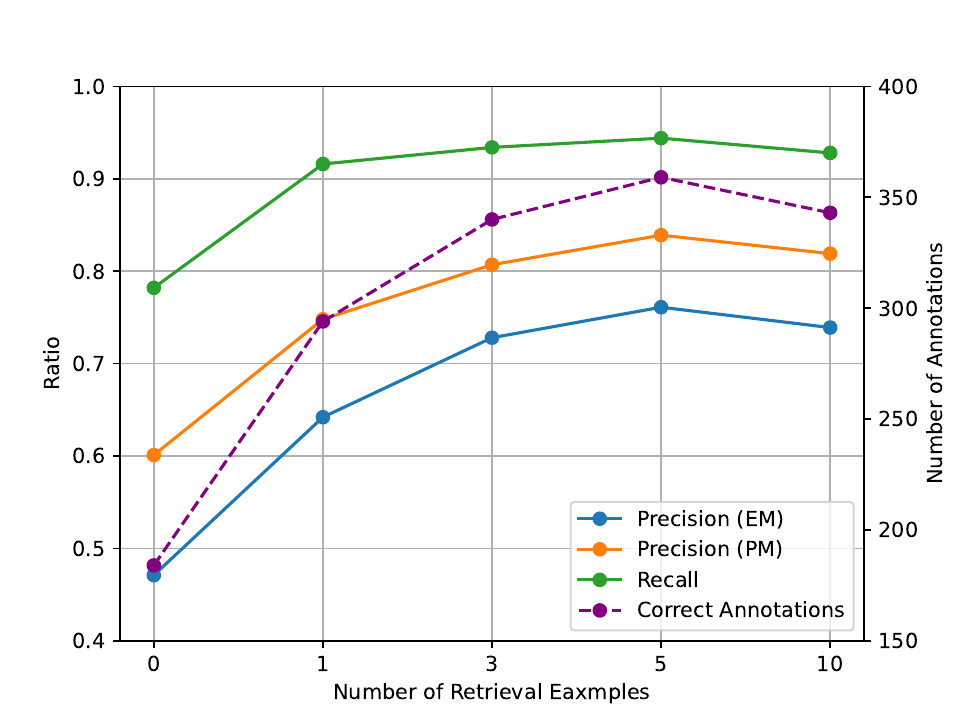}
    \caption{Performance with Varying Retrieved Examples}
    \label{fig:sensitive}
\end{figure}

One key parameter of \tool is the number of guiding examples $K$ provided for reasoning. 
In this RQ, we aim to investigate the optimal settings of the retrieved samples. 
To do so, we retrieved TOP-1, TOP-3, TOP-5, and TOP-10 samples for subsequent analysis.  
Figure~\ref{fig:sensitive} illustrates the performance of \tool regarding different numbers of retrieved examples.
\begin{itemize} [leftmargin=*]
    \item The overall performance of \tool increases as $k$ increases. 
    For example, the Precision (EM) increased from 0.471 to 0.761 when we provided guiding examples from 0 to 5. 
    This shows that more guiding examples can provide more contextual information to help LLMs better understand the project-specific type updating. 
    \item However, adding more retrieved samples does not always bring performance improvement. 
    When we added retrieved samples to the Top-10 examples, the performance of \tool slightly decreased. 
    Providing too many guiding samples can bring in more noise and incur bigger challenges for LLMs with information overloading problems. 
    Furthermore, the maximum context length of the large language model also limits the number of retrieved examples, leading to potential failure cases.
\end{itemize}

\find{
{\bf Answer to RQ3:} We recommend setting the retrieved examples to 5 when using \tool in practice. 
}

\subsection{Manual Analysis}
\begin{figure}
    \centering
    \includegraphics[width=0.48\textwidth]{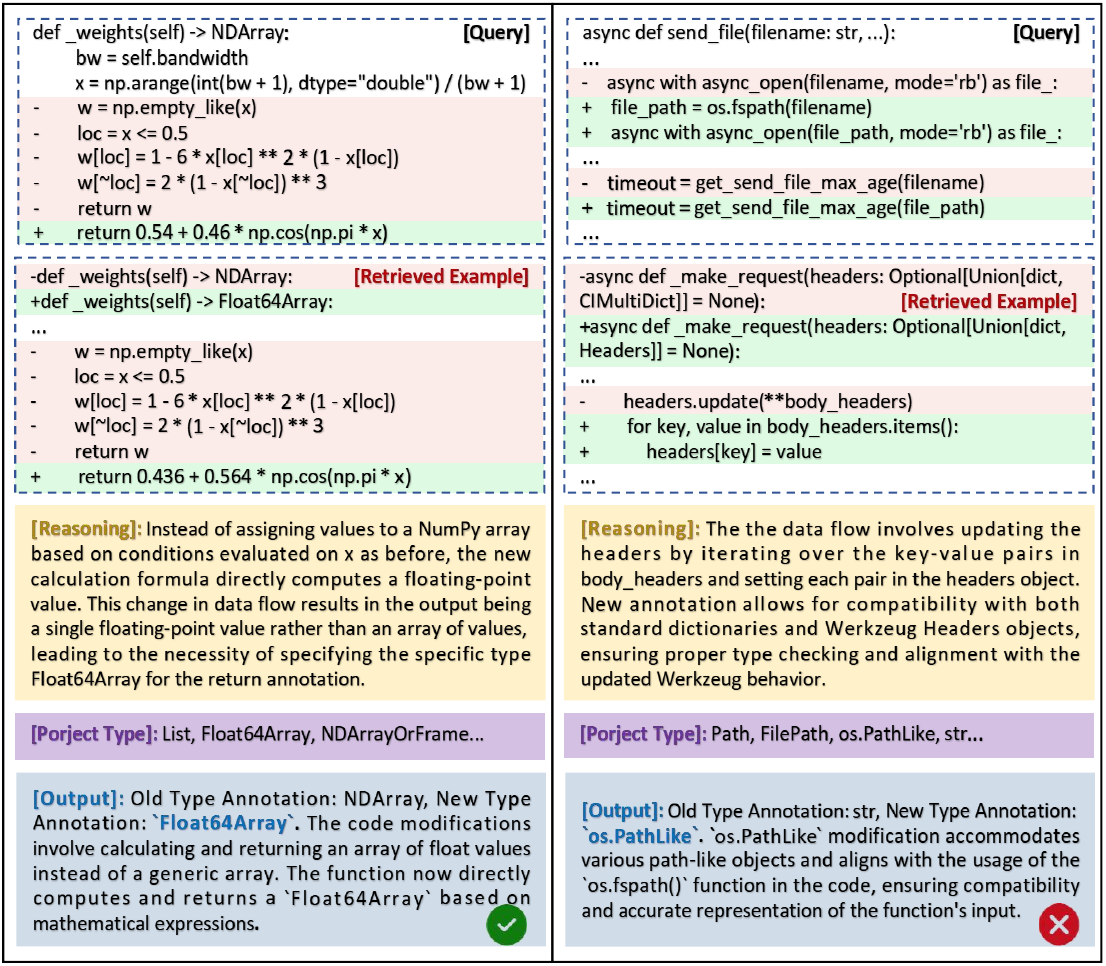}
    \caption{Updating Examples of \tool}
    \label{fig:discuss}
\end{figure}

To better illustrate the abilities of \tool, we manually inspected the test results, including the successful samples and failed samples. 
Based on our inspection, we summarize the advantages and limitations of our approach, Figure~\ref{fig:discuss} demonstrates a successful case (left) and a failed case (right) of our approach. 
\begin{enumerate}[leftmargin=*]
    \item \textit{Why does \tool works?} \tool has its advantages over baselines in terms of two aspects: firstly, we utilize LLMs for capturing semantic relationships among code-type co-changes. 
    Secondly, we augment LLMs with project-specific contexts (e.g., guiding examples and annotation candidates) for this task. 
    The left case of Figure~\ref{fig:discuss} illustrates an example correctly predicted by \tool, which updated the type annotation from \texttt{NDArray} to \texttt{Float64Array}. \tool first retrieved an example that performs a similar code modification in another class. Based on the retrieved example, the \textbf{\textsc{Reasoning Agent}} identified the data flow change of the code modification and explained how this change affects the data type of the return value. 
    Subsequently, the \textbf{\textsc{Retrieval Agent}} successfully extracted the \texttt{Float64Array} as an annotation candidate, as defined by the developer. Augmented with this information, \tool accurately updated this type of annotation. 
    In contrast, TypeGen predicts the type annotation as \texttt{Float64}, because it lacks the context information of the code change, which hindered its ability to accurately infer the type from the source code alone. 
    \item \textit{Why does \tool fails?} A common bad situation is that the code changes are too complicated for \tool to handle perfectly. 
    Under such circumstances, \tool may not be able to correctly reason the code-type co-changes. 
    Another bad situation is that if the code change pattern rarely occurs in the dataset, \tool may not be able to capture and apply it without any clues. 
    The right side of Figure~\ref{fig:discuss} illustrates an example in which \tool fails to predict correctly. 
    Despite the \textbf{\textsc{Retrieval Agent}} provided the correct \texttt{FilePath} as an annotation candidate, \tool mistakenly updated this type annotation as \texttt{PathLike} instead of the user-defined \texttt{FilePath}. 
    This is because the code changes or guiding examples do not provide enough information for \tool to infer correctly. 
    These examples reveal that our model’s performance may decline when the retrieval module cannot provide meaningful guidance, such as in newly created projects, projects with poorly maintained type annotations, or those with scarce historical type changes. 
\end{enumerate}

\find{\textbf{Manual Analysis Result:} Our approach benefits from enhancing LLMs with project-specific knowledge, but may fail where the retrieve module cannot provide effective guidence.}




\section{Practical Application}
\label{sec:wild}
Our ultimate goal is to help developers update the type annotations just-in-time after they change the code. 
To do this, we further conduct a practical application by updating the outdated type annotations in real GitHub projects.  

\subsection{Experimental Setup}

To perform type annotation updating, we first need to collect outdated type annotations from GitHub. 
To do so, we randomly selected ten popular well-annotated, open-source projects with more than 2k stars. 
These projects, which span various domains such as deep learning and project building, are listed in Table~\ref{tab:wild}. 
Outdated type annotations often lead to incompatibilities and can be detected and reported by static type checkers, we employed Pyre~\cite{pyre}, a widely used type checker, to identify these outdated annotations. 
\xzp{Since these incompatibilities, caused by outdated type annotations, frequently emerge after code changes, we traversed all commits and compared the reported incompatible type warnings between the pre-change and post-change versions.}

Particularly, we recorded all annotated code elements and their annotation types, and the function signature of each incompatible type warning. 
We regard the warnings in both the pre-change and post-change versions with the same recorded information as the same warning, which could not be attributed to outdated type annotations. 
Furthermore, we also ignored incompatible type warnings whose function signatures were absent in the pre-commit version, indicating they were caused by newly added type annotations. 
\xzp{Finally, we manually reviewed the remaining incompatible type warnings to determine whether they were caused by code changes and whether they still existed in the latest version of the project.}

\xzp{As a result, we identified 25 incompatibilities caused by outdated type annotations across 10 GitHub projects, all of which persisted until we submitted pull requests to resolve them.}
Following that, we checked out the commit introducing the outdated type annotation, we then extracted the code change and old annotation type from this commit and fed them into our approach. 
Finally, \tool can generate a new type annotation to replace the old one. 

\subsection{Practical Evaluation Results}
After generating type annotations for all 25 outdated cases by \tool, we submitted the updated type annotations as pull requests to the original GitHub repository to get developers' feedback. 
To avoid subjective bias, the developers do not know that the pull request results are automatically generated by our tool. 
\textbf{We submitted 25 pull requests to 10 software projects, 20 of them have already been confirmed by project developers or maintainers, and 5 of them have been rejected.}
The final results of our practical application are illustrated in Table~\ref{tab:wild}.

\begin{table}
  \centering
  \caption{Practical Application Result}
  \label{tab:wild}
  \begin{tabular}{lcccc}
    \toprule
    Project  & Merged & Changed& Rejected & Submitted\\
        \midrule
    Sktime &3 & 0 & 3&6\\
    Vyper & 1 & 0 & 0 & 1\\
    Integration & 0 & 0& 1 & 1\\
    Modin & 1 & 0& 0 & 1\\
    Akshare & 2 & 0 & 0 & 2\\
    Black & 0 & 0& 1 & 1\\
    Raster-vision & 5 & 0 & 0 & 5\\
    Prefect & 4 & 0 & 0& 4\\
    Electricitymaps & 0 & 2 & 0& 2\\
    Meson & 2 & 0 & 0& 2\\
    \midrule
    Total & 18 & 2 & 5& 25\\

    \bottomrule
  \end{tabular}
\end{table}

\begin{figure}
    \centering
    \includegraphics[width=0.45\textwidth]{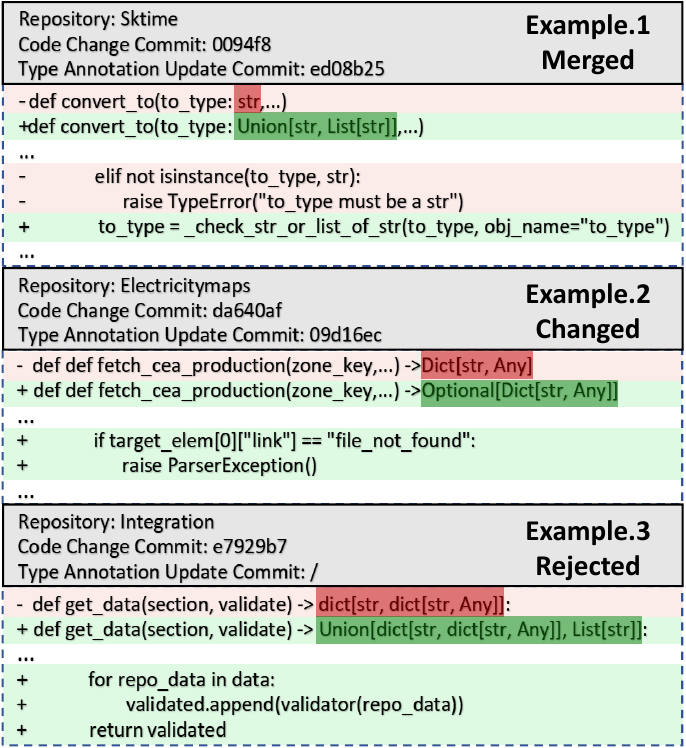}
    \caption{Examples of Submitted Pull Request}
    \label{fig:wild}
\end{figure}

We found that most of the type annotation updating pull requests (72\%) are directly approved and merged by developers. 
Figure~\ref{fig:wild}. 
Ex.1 illustrated an example of a merged pull request in project \texttt{Sktime}, \tool successfully updated the type annotation of the argument \texttt{to\_type} from \texttt{str} to \texttt{Union[str, List[str]]}. 
The developer of \texttt{Sktime} confirmed our pull request, replying that \textit{``As the support for lists of str was added after the original type hints, and these were forgotten to be updated.''} 
This confirmation validates the need for a tool to update type annotations in practice. 
Additionally, a developer of \texttt{Vyper} expressed interest in our tool and type annotation fixes, asking, \textit{``Interesting, thanks! Are you performing large scale type hint fixes?''}. 
Based on the high ratio of confirmed pull requests and the positive feedback from developers, we conclude that our tool can effectively and practically help developers handle the type annotation updating task, alleviating the burden of the code reviewing process for developers. 

Not all the updated type annotations were directly merged by developers. Two updated type annotations in \texttt{Electricitymaps} required changes. 
Figure~\ref{fig:wild}. Ex.2 illustrated one of the changed pull requests. 
\tool updated the return type annotation from return \texttt{Dict[str, Any]} to \texttt{Optional[Dict[str, Any]]}.
While the developer requested, ``\textit{Could you update the type annotation to match the expected format reported by ruff?}'' 
We modified the updated type annotation from \texttt{Optional[Dict[str, Any]]} to \texttt{Dict[str, Any]|None}, which retains the same semantics. 
According to PEP-604~\cite{PEP604}, \texttt{Optional[Type]} can be replaced by \texttt{Type|None} since Python 3.10. 
However, this version is higher than the Python version used by most projects in our codebase, making it challenging for \tool to predict type annotations that match new proposals. 
In the future, we plan to continuously expand our type annotation updating knowledge base to accommodate new proposals. 

Developers rejected five PRs generated by \tool. 
Figure~\ref{fig:wild} Ex.3 shows a rejected example. 
\tool predicted the updated return type annotation as \texttt{Union[dict[str, dict[str, Any]], List[str]]}, while the developer of \texttt{Integration} rejected the pull request by replying \textit{``\textit{None of the data paths contains a list of strings.}''} 
In the case of \texttt{Black}, \tool updated a type annotation in a unit test, but the developer rejected this change, stating, \textit{``The test cases are meant purely for formatting; they are not meaningful Python code and should not be type checked.''} 
Specifically, \tool updated the annotations of function arguments whose default values were newly set to \texttt{None} in the code modifications by adding the \texttt{Optional} type. While the developer acknowledged this update as technically correct, they preferred to omit it to maintain code readability. They explained, \textit{``In this case, it is considered 'obvious' that the arg is \texttt{Optional}, and adding this just lowers readability, so I would prefer not having \texttt{Optional} if the default is \texttt{None}.''}
Our approach is not perfect, for such cases, we still need developers to double-check the results. 

\vspace{5pt}
\find{\textbf{Practical Application Summary:} Among 25 outdated type annotations updated by \tool, 20 of them are confirmed by developers, which justifies the practical value of our tool.}

\section{Discussion}
In this work, we regard the output from the \textsc{\textbf{Updating Agent}} as the final result of \tool. 
Although \tool outperforms all baseline methods, it may still generate incorrect results (e.g., 5 generated annotations were rejected by developers in the practical application). 
In this section, we further explore the possiblity of adding a static analysis tool as a post-processing component to validate the generated results.

\noindent{\textbf{Experimental Setup.}}
To conduct the experiment, we leverage Pyre~\cite{pyre}, a widely used static analysis tool, as the annotation validator. 
Specifically, if a generated annotation passes Pyre's check, we accept it as the final output. 
Otherwise, if Pyre rejects the annotation, we use it as feedback and query the LLM to regenerate the annotation. 
This process is repeated up to five times. 
If the generated annotation still fails to pass Pyre’s validation after five attempts, we conclude that our approach is unable to produce the correct result for this case. 

\noindent{\textbf{Experimental Result.}}
The evaluation result is shown in Table~\ref{tab:validation}, where \tool+Val represents adding Pyre after our approach as a validator. 
We observe that: (1) with Pyre validation, both Precision (EM) and Precision (PM) improve by 10.4\% and 9.1\%, respectively, while Recall decreases from 0.944 to 0.776, leading to a drop in the final number of Correct Annotations. 
(2) Further analysis reveals that some correct updated annotations are rejected by Pyre, whereas some incorrect annotations are accepted. We attribute this phenomenon to the complex project contexts, particularly involving user-defined annotations, which pose significant challenges for static analysis tools.
(3) These results demonstrate that our \tool framework can be effectively integrated with existing static analysis tools for enhanced type validation. 
However, the gain in precision comes at the cost of reduced coverage. 
Exploring how to combine LLMs with static analysis tools to achieve optimal performance represents a promising research direction, but it is beyond the scope of this work. 

\find{\textbf{Discussion Summary:} \tool can be effectively integrated with static analysis tools for post type validation, but how to balance precision and recall remains an open research problem.}

\begin{table}
  \centering
  \caption{Integrated Validator Effectiveness Evaluation}
  \label{tab:validation}
    \resizebox{0.47\textwidth}{!}{
  \begin{tabular}{lcccc}
    \toprule
    Approach & Precision (EM) & Precision (PM) & Recall& \makecell[c]{Correct\\Annotations}\\
    \midrule
    \tool+Val & \textbf{0.840} & \textbf{0.915} & 0.776 & 326\\
    \tool & 0.761& 0.839 & \textbf{0.944} & \textbf{359}\\
    \bottomrule
  \end{tabular}
  }
\end{table}



\subsection{Threats to Validity}
\noindent{\textit{External Validity.}} 
The external validity relates to the generalization of our dataset. 
Our dataset is built from Python projects and only contains the updates of method argument and return type annotations, which may not be representative of all dynamic programming languages and types of annotations. 
However, our proposed model is independent of programming languages and type annotation types, it can be easily applied to projects in other languages and update other types of annotations. 

\noindent{\textit{Construct Validity.}} 
The construct validity concerns the relation between theory and observation. 
We collected code-type co-change samples from the same commit due to challenges in identifying the fix commit for outdated type annotations. 
This construction method differs from our expected usage scenario, where updated type annotations may be forgotten by developers. 
To validate the effectiveness of \tool in assisting developers with these overlooked outdated type annotations, we conducted a practical application study with real-world projects. 

\noindent{\textit{Model Validity.}} The model validity relates to the model selection that could affect the performance of our approach. 
Given that the GPT-3.5 model has been pre-trained on a vast corpus, including code from GitHub, this raises potential concerns about data leakage between our constructed dataset and LLM's training data. 
Nonetheless, the significant enhancements achieved by \tool over the baseline TypeGen, which is also implemented by GPT-3.5 model, demonstrate that \tool's effectiveness is not merely a result of data memorization.

\section{Related Work}
In this section, we describe the related studies on type inference and type error fixing. 
\subsection{Type Inference}
Due to the importance of type annotation in Python, previous studies have attempted to automatically infer type annotations from the code. These approaches can be categorized into two types: rule-based methods and learning-based methods. Rule-based methods predict types based on predefined rules. They typically generate and resolve type constraints from the source code through static analysis~\cite{furr2009profile, hassan2018maxsmt, hu2021static}, dynamic analysis~\cite{an2011dynamic, ren2013ruby}, or probabilistic rules~\cite{kazerounian2020sound, xu2016python}. Tools like pytype~\cite{pytype}, Pysonar2~\cite{pysonar2}, and Pyre~\cite{pyre} are designed to be accurate and perform well for simple built-in and generic types. However, the dynamic nature of programming languages~\cite{richards2010analysis} makes it difficult for these tools to handle user-defined types and more complex generic types effectively.
The learning-based approach involves training a deep learning model on a dataset of collected type annotations, such as ManyTypes4Py~\cite{mt4py2021}. 
Given a query code snippet, the model predicts the most likely type based on the proximity of high-dimensional vectors. Some studies treat type inference as a classification problem~\cite{peng2022static, mir2022type4py, yan2023dlinfer, jesse2021learning, weilambdanet}. 
For instance, DLInfer~\cite{yan2023dlinfer} employs static analysis combined with a bidirectional gated recurrent unit (GRU) model to accurately infer variable types. 
TypeWriter~\cite{pradel2020typewriter} integrates probabilistic type prediction with search-based refinement to enhance the accuracy of type inference. However, these approaches are limited in their ability to predict user-defined types, which restricts their practical utility in real-world projects. Generation-based models are built upon large language models (LLMs). 
These LLMs, pre-trained on vast corpora, exhibit a strong general understanding and generation capability. TypeT5~\cite{weitypet5} fine-tunes the CodeT5~\cite{wang2021codet5} model to enhance type inference accuracy. TypeGen~\cite{peng2023generative} constructs a chain-of-thought (CoT) prompt to guide the GPT-3.5~\cite{Chatgpt} model in inferring types. TIGER~\cite{wang2024tiger} designed a two-stage generating-then-ranking framework to infer type annotation by fine-tuning pre-trained code models.
In this study, our focus is on the task of updating type annotations, and we enhance LLMs with retrieval-augmented generation (RAG) technology to effectively handle the high frequency of user-defined type changes.

\subsection{Type Error Fixing}
Since type error is popular in Python, which makes up more than 30\% of all defects~\cite{khan2021empirical}, recently researchers have attempted to automatically repair type errors~\cite{10.1145/3540250.3549130, peng2024domain, chow2024pyty}. 
For example, PyTER~\cite{10.1145/3540250.3549130} repairs the type error based on a series of pre-defined rules. 
Peng et al.~\cite{peng2024domain} fix such type errors via leveraging LLMs by improving prompts with fix templates. Pyty~\cite{chow2024pyty} mitigates type errors with a fine-tuned LLM and verifies the fix patches by static checkers. 
Moreover, the automatic program repairing tools are also capable of fixing type errors.
Xia et al.~\cite{xia2023automated} leverages the LLMs to repair the programs automatically by provide the model with similar fix examples.
Li et al.\cite{li2024hybrid} supports the LLMs by the static analysis result and achieve significant performance in program repairing.
InferFix~\cite{jin2023inferfix} use LLMs to repair the errors which detected by the static analysis tool (i.e., Infer).
These previous studies assume the correctness of type annotation and focus on fixing the type errors from source code aspects. 
In this study, we found the problem of outdated type annotations and try to fix the type errors related to type annotations, we aim to update the obsolete type annotations just-in-time which are outdated after code changes.

\section{Conclusion}
In this paper, we propose a novel task, Just-In-Time Type annotation
updating, and we develop a tool, named \tool, to automate this task. \tool is based on Large Language Models and introduces customized RAG techniques to effectively handle the characteristics of this task. We build the first Python type annotation updating dataset which includes 36,796 code-type co-change samples. According to the evaluation, \tool surpasses the state-of-the-art type
inference model (i.e., TypeGen) by 41.9\%. Moreover, 20 out of 25 type annotation updates generated by our approach have already been confirmed by developers, showing our approach’s potential for increasing software quality and maintainability.

\begin{acks}
This research is supported by the National Science Foundation of China (No. 62572322).  
This research is partially sponsored by the Shanghai Sailing Program (23YF1446900) and the CCF-Tencent Rhino-Bird Open Research Fund. 
We also thank the anonymous reviewers for their insightful comments and suggestions. 
\end{acks}

\balance
\bibliographystyle{ACM-Reference-Format}
\bibliography{main}

\end{document}